\documentclass[aps,prl,epsfigure,showpacs,twocolumn]{revtex4}
\usepackage{graphicx}
\usepackage{amsmath}
\usepackage{amstext}
\usepackage{latexsym}
\usepackage{psfig}
\usepackage{amsfonts}
\usepackage{bm}
\usepackage{amssymb}
\usepackage{amscd}
\begin{document}
\title{Security against the Invisible Photon Attack for the Quantum Key Distribution with Blind Polarization Bases}
\author{Won-Ho Kye}
\affiliation{The Korean Intellectual Property Office, Daejeon
302-701, Korea}
\author{M. S. Kim}
\affiliation{School of Mathematics and Physics, Queen's University, Belfast BT7 1NN, United Kingdom}
\date{\today}
\begin{abstract}
In this paper, we briefly show how the quantum key distribution
with blind polarization bases [Kye et al., Phys. Rev. Lett. 95,
040501 (2005)] can be made secure against the invisible photon
attack.
\end{abstract}
\maketitle
Recently, we introduced a new quantum key distribution
(QKD)\cite{Kye}-we call this as the KKKP to stand for initials of
the authors-using random polarization bases and three-way
communication between Alice and Bob, two legitimate users of the
key. In the KKKP, Alice chooses a random value of angle $\theta$
and prepares a qubit pulse with the polarization of that angle.
Upon reception of the qubit, Bob chooses another random value of
angle $\phi$ and further rotates the polarization direction of
the received photon state by $\phi$ and then returns  to Alice.
Alice encodes the message by rotating the polarization angle by
$\pm \pi/4$ after compensating the angle by $-\theta$. Bob reads
the photon state by measuring the polarization, after compensating
the angle by $-\phi$.  Alice and Bob shall choose random angles,
$\theta$ and $\phi$, for each transmission of qubits. This will be
continued until the desired number of bits are created.  We
extended it to embody a qubit by a set of two pulses in order to
make the scheme robust against the impersonation attack. In
principle, the KKKP can be used for a secure direct communication
channel even though this is not what is claimed by the authors
because of the random security check.

In a recent paper\cite{Cai}, Cai questioned the security of the
so-called ping-pong quantum communications under invisible photon
attack (IPA). Here, a ping-pong quantum communication is a way of
secure direct communication based on quantum theory pioneered by
Bostr\"om and Felbinger\cite{Bostrom}.  In fact, the IPA is a
type of the Trojan Horse attack\cite{Gisin} under which most of
the QKD's are insecure.

However, the KKKP is designed to be relatively strong against a
simple IPA as legitimate users' operations are all random. As Cai
pointed out, the randomness of the operations makes it impossible
to read the key only by sending and reading spy photons. The
eavesdropper (Eve) has to take at least one photon out from the
first travel of the qubit pulse from Alice to Bob then she
compares the polarization of the disbound photon(s) with that of
the spy photons which she sends to Alice along with the returning
qubit pulse from Bob to Alice.  As we point out in
Ref.~\cite{Kye}, if Alice and Bob randomly check the intensities
of the qubit pulses, Eve's action of taking one photon out from
the Alice-to-Bob channel should be noticed. In practice, the
inefficiency of the photo detection may be a problem and in
particular for a lossy channel, it is impossible to distinguish
Eve's action from the channel loss.

It is, however, important to note another advantage of the KKKP
due to the fact that the returning pulse from Bob to Alice has
another random parameter $\phi$ given by Bob. Eve should be able
to separate her spy photons from the legitimate photons to read
Alice's action of $-\theta\pm\pi/4$. If Eve's spy photon is
indistinguishable then the photons, which travel finally to Bob
from Alice, will have the polarization angle as a mixture of
$\phi\pm\pi/4$ and $-\theta+\phi\pm\pi/4$.  Here, the second
polarization angle is due to Eve's spy photons which are assumed
initially to have 0 polarization angle. Now, we know the reason
why Eve's spy photons should be distinguishable from the
legitimate photons.  As Cai said, one way to achieve this is to
use photons of different frequencies.  In order to prevent from
such the attack, Alice and Bob should choose the optical devices
which have a very small frequency bandwidth.  These days, the
bandwidth of optical devices is as narrow as 0.1 or 0.01nm which
is comparable to the laser linewidth so that Eve's attack should
be able to be securely defended.  An optical grating to filter out
unwanted frequencies may be used in combination with such the
narrow bandwidth devices.

In this short communication, the security of the KKKP has been
reviewed.

\acknowledgments {\it Acknowledgments}- We thank the UK
Engineering and Physical Science Research Council.


\begin{thebibliography}{150}
\bibitem{Kye} W.-H. Kye, C. Kim, M. S. Kim and Y.-J. Park, \prl
{\bf 95}, 040501 (2005).
\bibitem{Cai} Q-Y Cai, quant-ph/0508002 (2005).
\bibitem{Bostrom} K. Bostr\"om and T. Felbinger, \prl {\bf 89},
187902 (2005).
\bibitem{Gisin} N. Gisin, S. Fasel, B. Kraus, H. Zbinden and G.
Ribordy, quant-ph/0507063 (2005).

\end{thebibliography}
\end{document}